# Median Mishaps between Chirality and Spin-Orbit Torques via Asymmetric Hysteresis

Minhwan Kim[1,2] and Duck-Ho Kim[1*]

[1]Center for Spintronics, Korea Institute of Science and Technology, Seoul, 02792, Republic of Korea

[2]Department of Physics and Astronomy, Seoul National University, Seoul, 08826, Republic of Korea

[*]e-mail: uzes@kist.re.kr (D.-H. Kim)

# ABSTRACT

Averaged observations of physical phenomena have been utilized for comprehending specific occurrences in nature; however, these may overlook the crucial characteristics of individual events, thereby leading to diverse conclusions. For example, individual wave occurrences such as superposition and interference yield markedly divergent outcomes when viewed in detail. Similarly, this study enhances the comprehensive framework of spin-orbit torque (SOT) within the hysteresis loop shift measurement by employing the average of effective magnetic fields arising from two distinct magnetic reversals. This approach facilitates the presentation of a physically descriptive SOT model, previously characterized only by single chirality qualitatively. By integrating this model with established measurement methodologies and theoretical paradigms, we advance a theoretical framework based on the magnetic domain-wall chirality of individual polarizations and aim to elucidate the phenomena of SOT with clarity. The anticipated outcomes include the rectification of inaccuracies in widely employed measurement methodologies and the enhancement of our comprehension of the fundamental physics, which are expected to propel advancements in next-generation spintronics materials and devices.

# I. INTRODUCTION

The manipulation of magnetic states induced by electric current represents a pivotal aspect for the development of future spintronics devices[1–3]. Recently, spin-orbit torques (SOTs) have been observed to arise from various physical phenomena, including the spin Hall[4-10], Rashba–Edelstein[11–19], and orbital Hall effects[20–24], wherein the underlying principle is associated with the charge-to-spin conversion, inducing effective torques on magnets[4–24]. The advancement of these applications and the exploration of new physical phenomena is heavily reliant on the development and accurate measurement of SOT. Numerous techniques for SOT measurement have emerged, which often center on the detection of an effective magnetic field arising from two distinct phenomena: 1) oscillation of magnetization from the SOTs[25–28] and 2) the interaction between the magnetization of the domain wall (DW), chirality of the DW, and the flow of spin current[29–32].

The commonly adopted technique for assessing SOT efficiency and its characteristics involves the utilization of current-induced hysteresis loop shift, which is commonly believed to be closely related to the DW chirality[32–38]. Previous investigations have attempted to understand SOT efficiency by employing a qualitative physical model or an empirical notion relating DW chirality to the in-plane magnetic field[32,36]. Consequently, it has become imperative to propose a comprehensive physics model, this is supported by both experimental and theoretical demonstrations to elucidate this phenomenon. The ambiguity stemming from this circumstance result in inaccurate assessments of the essential physical parameter termed the Dzyaloshinskii-Moriya interaction (DMI)[39,40], which is pivotal for understanding the chiral spin structure[15,16,41–44]. Consequently, it fosters misconceptions within the underlying physics and results in inaccurate understanding of spintronics device characteristics, thereby posing significant obstacles to the progress of device development[1–3].

In this study, we propose a physics model for SOT measurement based on the current-induced hysteresis loop shift, aiming to obtain a comprehensive understanding of the SOT phenomena both experimentally and theoretically (Fig. 1). The analytical framework was established by averaging the DW chirality for individual polarizations and validated through experiments conducted on magnetic films with diverse physical attributes. This resolves previous misunderstandings that focused on the relationship between chirality and SOT observed in loop shift experiments. In experimental settings, the in-plane magnetic field versus SOT efficiency from loop shift results reflects an average SOT effect from two different chiralities. Therefore, simple chirality cannot adequately explain SOT characteristics; this approach yields an averaged value that includes inherent errors from each chirality on SOT (see the schematic illustration in Fig. 1). Consequently, our work identifies this as a fundamental misunderstanding of the averaged relationship between chirality and SOT.

## II. THEORETICAL MODEL

To elucidate the efficiency of SOT concerning the in-plane magnetic field $\mu_0 H_x$, as inferred from hysteresis loop shift measurements[32–38], we initially outlined the scenario of the physical model in accordance with the specified experimental protocol[32,36]. When accessing hysteresis loops for ferromagnetic materials, the switching field is intricately correlated with the stochastic nucleation of magnetic domains, which is influenced by factors such as local defects or etc.[45–48]. However, when investigating the SOT efficiency under $\mu_0 H_x$ based on hysteresis loop shift measurements, we employed a simplified assumption that the initial nucleation position remained fixed, disregarding any randomness originating from the film. Under these controlled conditions, the hysteresis loop shift measurement can be conceptualized as a process involving the domain nucleation and magnetization switching with different polarities while sweeping the magnetic field.

Figure 2**A** illustrates the schematics of the proposed model. As the out-of-plane magnetic field $\mu_0 H_z$ traverses from the negative to positive direction, an upward domain originates from a nucleation point (indicated by the yellow dot). Subsequently magnetization switching occurs at certain positive magnetic field, which is related to the transition from the DW creep motion to DW depinning motion[49–51] as depicted in Fig. 2**A**. In Fig. 2**B**, the transition field denotes the magnetization switching field from the up-down DW movement $\mu_0 H_{SW}^+$ (indicated by the red solid line). A similar trend is observed for the downward domain as the magnetic field sweeps from the positive to negative direction, and vice versa where the switching field of the down-up DW is denoted as $\mu_0 H_{SW}^-$ (see the blue solid line in Fig. 2**B**). In the absence of current, as $\mu_0 H_{SW}^+ = \mu_0 H_{SW}^{0+}$ and $\mu_0 H_{SW}^- = \mu_0 H_{SW}^{0-}$ exhibit a purely symmetric effect, $\mu_0 H_{SW}^{0+}$ unequivocally equals $-\mu_0 H_{SW}^{0-}$. In the presence of an electric current, SOT functions as an effective magnetic field $\mu_0 H_{SOT}^\pm$ (where + and – correspond to the up-down and down-up DW polarities, respectively) resulting in a shift in the switching fields: $\mu_0 H_{SW}^+ = \mu_0 H_{SW}^{0+} + \mu_0 H_{SOT}^+$ and $\mu_0 H_{SW}^- = \mu_0 H_{SW}^{0-} + \mu_0 H_{SOT}^-$.

In this scenario, the effective magnetic field resulting from SOT exerts an influence on the center field of the hysteresis loop, denoted as $\mu_0 H_{CF}$, which represents the average of the switching fields for each polarity, expressed as $\mu_0 H_{CF} = (\mu_0 H_{SOT}^+ + \mu_0 H_{SOT}^-)/2$. The SOT efficiencies corresponding to up-down and down-up DWs are indicated as $\varepsilon^\pm$. By scaling $\mu_0 H_{SOT}^\pm$ with the current density $J$, we obtain $\varepsilon^\pm = \partial \mu_0 H_{SOT}^\pm / \partial J$. Notably, the DW chirality can be determined from $\varepsilon^\pm$ with respect to the in-plane magnetic field $\mu_0 H_x$, where a Néel-type DW is identified by the point at which $\varepsilon^\pm$ saturates to its maximum $\varepsilon_{sat}^\pm$, while a Bloch-type DW emerges when $\varepsilon^\pm$ equals 0, transitioning from Néel to Bloch otherwise. The SOT efficiency in the hysteresis loop shift measurement, denoted as $\varepsilon_{LS}$, can be expressed by the simple relationship $\varepsilon^{LS}_{\square}(\mu_0 H_x) \equiv [\varepsilon^{+}_{\square}(\mu_0 H_x) + \varepsilon^{-}_{\square}(\mu_0 H_x)]/2$, representing the average of the up-down and down-up DW's chiralities

To validate our proposed model, an analytical approach was employed. According to the SOT theory, the normalized SOT efficiency $\varepsilon^{\pm}_{norm}(\mu_0 H_x) \equiv \varepsilon^{\pm}/\varepsilon^{\pm}_{sat}$ can be transformed into the equilibrium azimuthal angle of the DW center magnetization $\psi^{\pm}_{eq}(\mu_0 H_x)$, where this transformation is simplified as $\varepsilon^{\pm}_{norm}(\mu_0 H_x) = \cos\psi^{\pm}_{eq}(\mu_0 H_x)$ (see the illustration of $\cos\psi^{+}_{eq}(\mu_0 H_x)$ in Fig. 2**C**)[14,52–55]. Utilizing a previously established theoretical DW model, the equilibrium azimuthal angle for each up-down and down-up DW, $\psi^{\pm}_{eq}(\mu_0 H_x)$, which is determined by minimizing the DW energy, can be expressed as,

$$\cos\psi^{\pm}_{eq}(\mu_0 H_x) = \begin{cases} \dfrac{\mu_0 H_x \pm \mu_0 H_{DMI}}{\mu_0 H_S} & \text{for } |\mu_0 H_x \pm \mu_0 H_{DMI}| < \mu_0 H_S \\ \pm 1 & \text{otherwire} \end{cases}, \quad (1)$$

where $\mu_0 H_{DMI}$ is the DMI induced effective magnetic field and $\mu_0 H_S$ is the DW anisotropy field, representing the magnitude of the in-plane magnetic field required to transform the Bloch-type DW to the Néel-type DW as indicated by the green arrow in Fig. 1**c**. Therefore, the normalized SOT efficiency in the hysteresis loop shift measurement, $\varepsilon_{norm,LS}(\mu_0 H_x)$, can be elucidated by $\cos\psi^{\pm}_{eq}(\mu_0 H_x)$, as follows:

$$\varepsilon^{LS}_{norm}(\mu_0 H_x) = \frac{[\cos\psi^{+}_{eq}(\mu_0 H_x) + \cos\psi^{-}_{eq}(\mu_0 H_x)]}{2}. \quad (2)$$

Notably, in all instances, due to its averaged characteristics, the $\varepsilon^{LS}_{norm} - \mu_0 H_x$ curve inherently exhibited symmetry, thus losing information about the DW chirality due to its averaged nature.

Figures 3**A**–**D** present the numerical calculations of $\varepsilon^{\pm}_{\text{norm}}(\mu_0 H_x)$ and $\varepsilon^{\text{LS}}_{\text{norm}}(\mu_0 H_x)$ derived using Eq. (2) for various magnetic properties of $|\mu_0 H_{\text{DMI}}|$ and $\mu_0 H_S$ : (**A**) $|\mu_0 H_{\text{DMI}}| > \mu_0 H_S$, (**B**) $|\mu_0 H_{\text{DMI}}| = \mu_0 H_S$, (**C**) $|\mu_0 H_{\text{DMI}}| < \mu_0 H_S$ and (**D**) $\mu_0 H_{\text{DMI}} = 0$ mT. Here, $\varepsilon^{+}_{\text{norm}}(\mu_0 H_x)$, $\varepsilon^{-}_{\text{norm}}(\mu_0 H_x)$, and $\varepsilon^{\text{LS}}_{\text{norm}}(\mu_0 H_x)$ are indicated by red, blue, and green solid lines, respectively. The results clearly depicted that $\varepsilon^{\pm}_{\text{norm}}(\mu_0 H_x)$ followed a step-like function according to Eq. (2), whereas the average value, $\varepsilon^{\text{LS}}_{\text{norm}}(\mu_0 H_x)$, exhibited a symmetric function passing through the origin. Moreover, the dependence of $\mu_0 H_x$ on $\varepsilon^{\pm}_{\text{norm}}$ was significantly influenced by the relative difference in values $|\mu_0 H_{\text{DMI}}|$ and $\mu_0 H_S$. This finding implies that our theoretical framework, in principle, can elucidate the SOT efficiency through the hysteresis loop shift measurement.

## III. EXPERIMENTAL VERIFICATION

In experimental verification, it is essential to ensure controlled nucleation sites. Typically, the presence of random nucleation sites originating during the production of magnetic thin films persists, which influences the coercive field of the hysteresis loop and introduces variability. This can result in inaccuracies in the measurement of the SOT efficiency via hysteresis loop shift, thus highlighting the importance of protocols for regulating these nucleation sites[32,36,45–48].

To experimentally validate our proposed idea, a perpendicularly magnetized thin film was prepared with a structure composed 5-nm Ta/2.5-nm Pt/0.8-nm Co/1.0-nm Pt/3.0-nm Ta/1.5-nm Pt deposited on an Si substrate with 100-nm $SiO_2$ via DC magnetron sputtering. Employing photolithography, we fabricated Hall-bar shaped devices with 5-μm width and 500-μm length wire to electrically detect the magnetic signal. The initiation of domain nucleation at the identical position was achieved via adjustments to the magnetic property of the wire by softly milling the local area (indicated by the yellow-colored box in inset of Fig. 4**A**).

The experimental procedure commenced with the application of an electric current density $J$ along the x-directional wire, accompanied by a constant $\mu_0 H_x$. Subsequently, anomalous Hall voltage detection was conducted by sweeping the out-of-plane magnetic field $\mu_0 H_z$ from the negative to positive direction and vice versa, followed by the identification of the switching fields. Figure 4**B** shows the hysteresis loop under $\mu_0 H_x = -351.6$ mT and $J = \pm 4.76 \times 10^{10}$ A/m². In Fig. 4**B**, the center field are denoted as $\mu_0 H_{\mathrm{CF}} \equiv (\mu_0 H_{\mathrm{SW}}^+ + \mu_0 H_{\mathrm{SW}}^-)/2$ (indicated by dashed lines). Clear shifts in the center field are observed depending on the sign of $J$. To determine the SOT effect, we investigate $\mu_0 H_{\mathrm{CF}}$ of the loop with respect to $J$, as shown in Fig. 4**C**, where the slope represents the SOT efficiency $\varepsilon_{\square}^{\mathrm{LS}}$. Figure 4**D** shows $\varepsilon_{\mathrm{norm}}^{\mathrm{LS}}$ as a function of $\mu_0 H_x$ (indicated by the green solid symbol). The result clearly indicated a tendency towards origin symmetry $\varepsilon_{\mathrm{norm}}^{\mathrm{LS}}$, as expected in Fig. 3.

To validate the concordance between the experimental results and our proposed theoretical framework, we conducted SOT efficiency measurements via DW depinning motion[29,30], thereby facilitating the determination of $\mu_0 H_{\mathrm{S}}$ and $\mu_0 H_{\mathrm{DMI}}$ [see Note. 1 in the Supplementary Information]. Using these parameters, we investigated the analytical results, derived from the measured $\mu_0 H_{\mathrm{S}}$ and $\mu_0 H_{\mathrm{DMI}}$ alongside Eq. (2), depicted by the yellow-green solid line in Fig. 4**D**. Notably, the experimental results obtained through hysteresis loop shift corroborate our proposed theoretical model.

A significant implication of our findings is that the configuration of the $\varepsilon_{\mathrm{norm}}^{\mathrm{LS}}(\mu_0 H_x)$ is influenced by two material parameters; $\mu_0 H_{\mathrm{S}}$ and $\mu_0 H_{\mathrm{DMI}}$. To substantiate this assertion, we examined the measured $\varepsilon_{\text{norm}}^{\text{LS}}(\mu_0 H_x)$ across a range of magnetic films, encompassing another values of $\mu_0 H_S$ and $\mu_0 H_{\text{DMI}}$ (asymmetric structure as Pt/Co/Pd for the case where $|\mu_0 H_{\text{DMI}}| > \mu_0 H_S$). Detailed sample structures with corresponding measurement values are presented in Table I. Figure 5 illustrates the relationship between $\varepsilon_{\text{norm}}^{\text{LS}}$ and $\mu_0 H_x$, where the green solid symbol represent experimental data points, depicted alongside the yellow-green solid line that represent analytical calculations derived using Eq. (2) and relevant material parameters (see Note. 1 in the Supplementary Information). Notably, a plateau is observed in the vicinity of $\mu_0 H_x = 0$ mT as expected in Fig. 3**A**, corresponding to the analytic calculation for the case where $|\mu_0 H_{\text{DMI}}| > \mu_0 H_S$. A significant alignment between experimental observations and theoretical predictions was observed across different sets of material parameters, thus providing compelling experimental evidence for the proposed model.

Our proposal holds significance on two fronts. Firstly, the SOT efficiency as a function of the in-plane magnetic field, which was assessed through hysteresis loop shift measurement, lacks intrinsic physical relevance to single DW chirality owing to its representation as an average of two distinct polarities of chiral DWs: up-down and down-up DWs. This resulted in loss and distortion of the transition dynamics between Bloch-type and Néel-type DWs. Despite previous reports attempting to characterize $\varepsilon_{\square}^{\text{LS}}(\mu_0 H_x)$ in terms of the DW chirality[32,36], it is unsuitable for characterization as it merely corresponds to the average of two distinct DW chirality. Secondly, within the context of SOT efficiency from hysteresis loop shift, the determination of $\mu_0 H_{\text{DMI}}$ is significantly influenced by the interplay between $\mu_0 H_S$ and $\mu_0 H_{\text{DMI}}$, rather than solely relying on the $\mu_0 H_x$ value at $\varepsilon_{\text{norm}}^{\text{LS}} = 1$. The criteria for determining $\mu_0 H_{\text{DMI}}$ from $\varepsilon_{\text{norm}}^{\text{LS}}(\mu_0 H_x)$ are summarized in Table II. This discrepancy results in significant experimental misinterpretation, owing to a critical error in the quantitative determination of DMI. Because of the complexity of characterizing $\varepsilon_{\square}^{\text{LS}}(\mu_0 H_x)$, we emphasize that focusing on one polarity of the switching field in the loop shift method is more straightforward for elucidating the interplay between chirality and SOT, which aligns with SOT efficiency measurements via DW depinning motion (see Note. 2 in the Supplementary Information) This approach avoids ambiguities and errors from averaging two DW polarities, thereby elucidating the DW chirality and ensuring accurate DMI determination.

## IV. CONCLUSION

This study proposed a theoretical framework for SOT measurement based on the current-induced hysteresis loop shift and experimentally demonstrated the rectification of significant inaccuracies in SOT observation by averaging effective magnetic fields from two distinct magnetic reversals. This underscored the tendency for average observations of SOT phenomena via hysteresis loop shifts to overlook the essential characteristics of individual chirality. These findings represent a substantial advancement in our understanding of the relationship between SOT and chiral DW phenomena in magnetic materials, while also providing a precise measurement platform. Therefore, this is expected to serve as a crucial driver in propelling the development of next-generation spintronics materials and devices.

TABLE I. Magnetic properties of various samples measured by $\varepsilon^+(\mu_0 H_x)$.

| Sample Structure [unit: nm] | $\mu_0 H_{\text{DMI}}$ [mT] | $\mu_0 H_S$ [mT] |
|---|---|---|
| Ta(5)/Pt(2.5)/Co(0.8)/Pt(1)/Ta(3)/Pt(1.5) | $-29.86 \pm 2.67$ | $71.25 \pm 8.79$ |
| Ta(5)/Pt(2.5)/Co(0.6)/Pd(3)/Pt(1.5) | $-232.42 \pm 2.19$ | $128.19 \pm 20.85$ |

TABLE II. Determination of $|\mu_0 H_{\text{DMI}}|$ from $\varepsilon^{\text{LS}}_{\text{norm}}(\mu_0 H_x)$. $|\mu_0 H_{\text{DMI}}|$ corresponding to $\mu_0 H_x$ which satisfies the given value of $\varepsilon^{\text{LS}}_{\text{norm}}$.

| | > | = | < | 0 |
|---|---|---|---|---|
| $\varepsilon^{\text{LS}}_{\text{norm}}$ | $\pm 1/2$ | $\pm 1/2$ | N/A | 0 |

**Acknowledgements**

We would like to thank Prof. Sug-Bong Choe for fruitful discussions and Editage ([www.editage.co.kr](www.editage.co.kr)) for English language editing. This work was supported by the National Research Foundation (NRF) of Korea (grant NRF-2022R1A2C2004493, 2N74520) and by the Korea Institute of Science and Technology (KIST) institutional program (grant 2E32951).

**Author Contributions**

D.H.K. conceived the idea and supervised the research. M.K. was responsible for conducting the experiments and fabricating the films and devices. M.K. and D.H.K. developed the theoretical framework and analyzed the data. The manuscript was written by M.K. and D.H.K. All authors participated in discussions of the results and reviewed of the manuscript.

**Additional information**

Correspondence and requests for materials should be addressed to D.H.K.

**Competing financial interests**

The authors declare no competing financial interests.

**Data availability**

All data are available in the main text or the Supplementary Information.

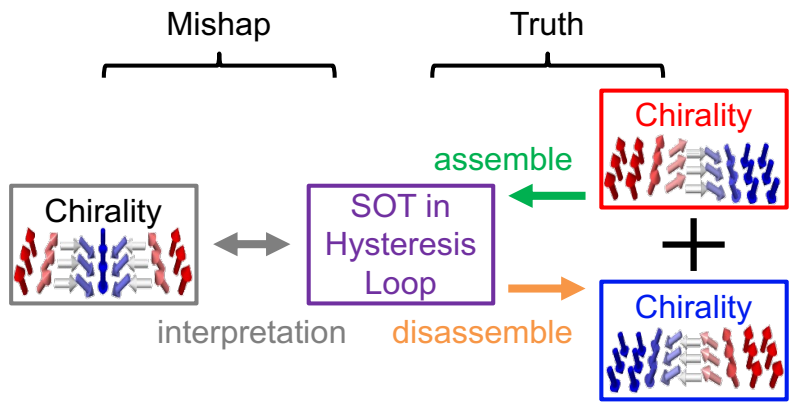

**Figure 1. Schematic Illustration of Median Mishap in the Asymmetric-Hysteresis-Induced SOT.** The left side represents the prevailing interpretation of asymmetric-hysteresis-induced SOT based on the single chirality, which can result in erroneous conclusions. The right side shows the assembly and disassembly of the key physics of SOT using two chralities, revealing the underlying physics of SOT within the hysteresis loop.

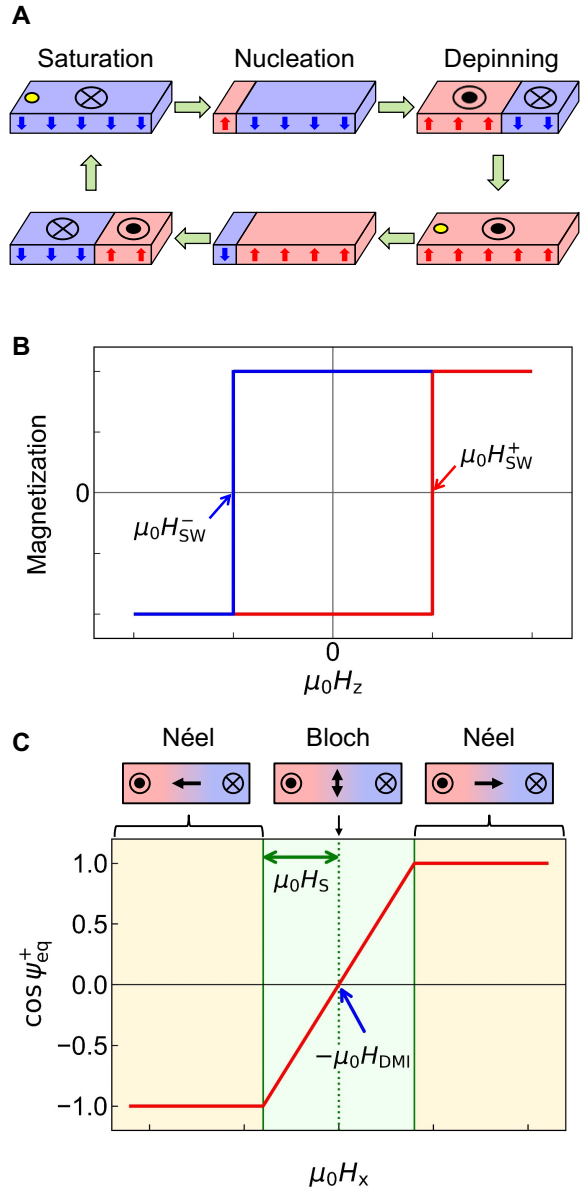

**Figure 2. Schematics of the Magnetic DW Chirality Transition. A**, schematics of the proposed model for domain nucleation. **B**, the magnetization switching field from the up-down DW and down-up DW movement. **C**, the equilibrium angle $\psi_{eq}$ as a function of $\mu_0 H_x$ and the DW types.

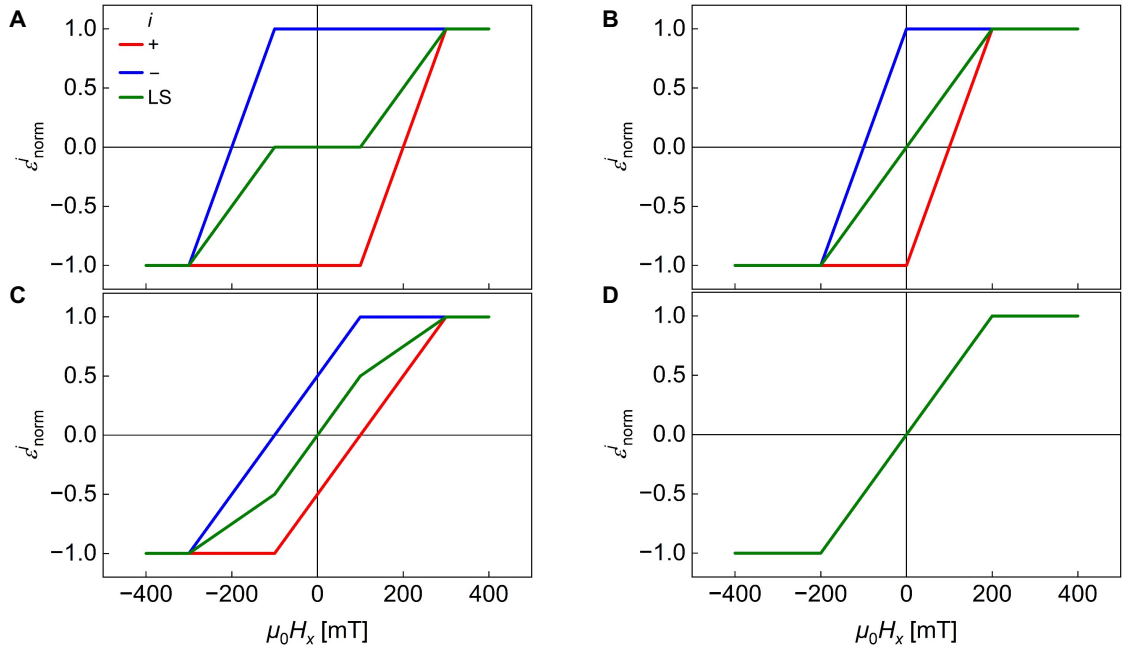

**Figure 3. Numerical Calculation of the Efficiency of SOT Concerning $\mu_0 H_x$ from Analytical Model based on Two Chiralities.** The schematics of the $\varepsilon^i(\mu_0 H_x)$, where $i = +$ dd $i = -$ and $i =$ LS correspond to up-down DW, down-up DW and the center of the hysteresis loop, respectively: **A**, $|\mu_0 H_{\text{DMI}}| > \mu_0 H_S$ case, where $\mu_0 H_{\text{DMI}} = -200$ mT and $\mu_0 H_S = 100$ mT. **B**, $|\mu_0 H_{\text{DMI}}| = \mu_0 H_S$ case, where $\mu_0 H_{\text{DMI}} = -100$ mT and $\mu_0 H_S = 100$ mT. **C**, $|\mu_0 H_{\text{DMI}}| < \mu_0 H_S$ case, where $\mu_0 H_{\text{DMI}} = -100$ mT and $\mu_0 H_S = 200$ mT. **D**, $\mu_0 H_{\text{DMI}} = 0$ mT case with $\mu_0 H_S = 200$ mT.

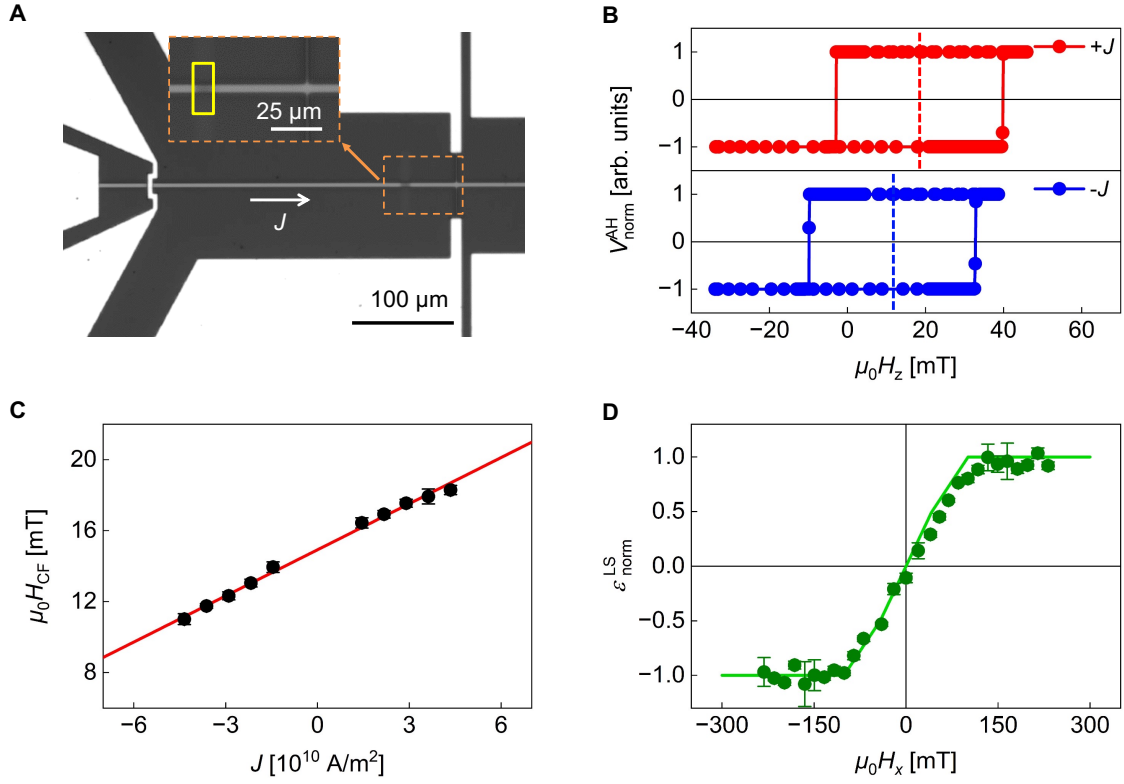

**Figure 4. Efficiency of SOT Concerning $\mu_0 H_x$ from Hysteresis Loop Shift Measurement for Pt/Co/Pt/Ta Sample. A**, schematics of the measurement setup with an optical image of the device. The anomalous Hall voltage was measured at the Hall bar. The inset magnifies the area within orange dashed box, highlighting the softly milled area marked by the yellow box. **B**, the hysteresis loop measured by anomalous Hall voltage under current density $J = 4.35 \times 10^{10}$ A/m² and a bias of $\mu_0 H_x = -198.2$ mT. The red and blue symbols represent the positive and negative current bias, with the current density of $J = 4.35 \times 10^{10}$ A/m². **C**, the plot of center field of the hysteresis loop $\mu_0 H_{CF}$ with respect to $J$. The red solid line indicates the linear best fit. **D**, the plot of $\varepsilon_{norm}^{LS}$ as a function of $\mu_0 H_x$ (experimental data is the green symbol). The yellow-green solid line denotes the analytical calculation from Eq. 1 using experimental parameters of $\mu_0 H_{DMI}$ and $\mu_0 H_S$.

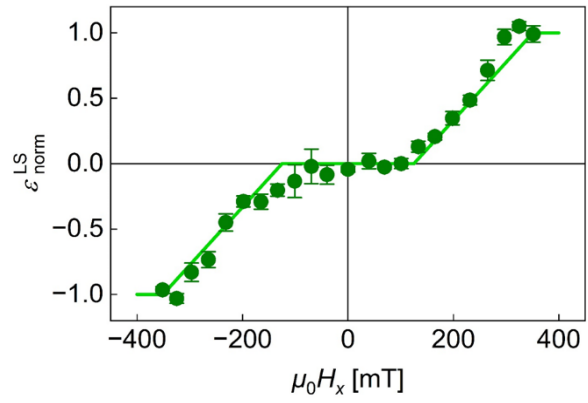

**Figure 5. Efficiency of SOT Concerning $\mu_0 H_x$ from Hysteresis Loop Shift Measurement for Pt/Co/Pd Sample.** The plot of $\varepsilon_{\text{norm}}^{\text{LS}}$ as a function of $\mu_0 H_x$ (experimental data is the green symbol). The yellow-green solid line denotes the analytical calculation from Eq. 1 using experimental parameters of $\mu_0 H_{\text{DMI}}$ and $\mu_0 H_{\text{S}}$.